# Observation of Gravitational Waves from a Binary Black Hole Merger

B. P. Abbott *et al.**

(LIGO Scientific Collaboration and Virgo Collaboration)

On September 14, 2015 at 09:50:45 UTC the two detectors of the Laser Interferometer Gravitational-Wave Observatory simultaneously observed a transient gravitational-wave signal. The signal sweeps upwards in frequency from 35 to 250 Hz with a peak gravitational-wave strain of $1.0 \times 10^{-21}$. It matches the waveform predicted by general relativity for the inspiral and merger of a pair of black holes and the ringdown of the resulting single black hole. The signal was observed with a matched-filter signal-to-noise ratio of 24 and a false alarm rate estimated to be less than 1 event per 203 000 years, equivalent to a significance greater than $5.1\sigma$. The source lies at a luminosity distance of $410^{+160}_{-180}$ Mpc corresponding to a redshift $z = 0.09^{+0.03}_{-0.04}$. In the source frame, the initial black hole masses are $36^{+5}_{-4} M_{\odot}$ and $29^{+4}_{-4} M_{\odot}$, and the final black hole mass is $62^{+4}_{-4} M_{\odot}$, with $3.0^{+0.5}_{-0.5} M_{\odot} c^2$ radiated in gravitational waves. All uncertainties define 90% credible intervals. These observations demonstrate the existence of binary stellar-mass black hole systems. This is the first direct detection of gravitational waves and the first observation of a binary black hole merger.

## I. INTRODUCTION

In 1916, the year after the final formulation of the field equations of general relativity, Albert Einstein predicted the existence of gravitational waves. He found that the linearized weak-field equations had wave solutions: transverse waves of spatial strain that travel at the speed of light, generated by time variations of the mass quadrupole moment of the source [1,2]. Einstein understood that gravitational-wave amplitudes would be remarkably small; moreover, until the Chapel Hill conference in 1957 there was significant debate about the physical reality of gravitational waves [3].

Also in 1916, Schwarzschild published a solution for the field equations [4] that was later understood to describe a black hole [5,6], and in 1963 Kerr generalized the solution to rotating black holes [7]. Starting in the 1970s theoretical work led to the understanding of black hole quasinormal modes [8–10], and in the 1990s higher-order post-Newtonian calculations [11] preceded extensive analytical studies of relativistic two-body dynamics [12,13]. These advances, together with numerical relativity breakthroughs in the past decade [14–16], have enabled modeling of binary black hole mergers and accurate predictions of their gravitational waveforms. While numerous black hole candidates have now been identified through electromagnetic observations [17–19], black hole mergers have not previously been observed.

The discovery of the binary pulsar system PSR B1913+16 by Hulse and Taylor [20] and subsequent observations of its energy loss by Taylor and Weisberg [21] demonstrated the existence of gravitational waves. This discovery, along with emerging astrophysical understanding [22], led to the recognition that direct observations of the amplitude and phase of gravitational waves would enable studies of additional relativistic systems and provide new tests of general relativity, especially in the dynamic strong-field regime.

Experiments to detect gravitational waves began with Weber and his resonant mass detectors in the 1960s [23], followed by an international network of cryogenic resonant detectors [24]. Interferometric detectors were first suggested in the early 1960s [25] and the 1970s [26]. A study of the noise and performance of such detectors [27], and further concepts to improve them [28], led to proposals for long-baseline broadband laser interferometers with the potential for significantly increased sensitivity [29–32]. By the early 2000s, a set of initial detectors was completed, including TAMA 300 in Japan, GEO 600 in Germany, the Laser Interferometer Gravitational-Wave Observatory (LIGO) in the United States, and Virgo in Italy. Combinations of these detectors made joint observations from 2002 through 2011, setting upper limits on a variety of gravitational-wave sources while evolving into a global network. In 2015, Advanced LIGO became the first of a significantly more sensitive network of advanced detectors to begin observations [33–36].

A century after the fundamental predictions of Einstein and Schwarzschild, we report the first direct detection of gravitational waves and the first direct observation of a binary black hole system merging to form a single black hole. Our observations provide unique access to the

*Full author list given at the end of the article.

properties of space-time in the strong-field, high-velocity regime and confirm predictions of general relativity for the nonlinear dynamics of highly disturbed black holes.

## II. OBSERVATION

On September 14, 2015 at 09:50:45 UTC, the LIGO Hanford, WA, and Livingston, LA, observatories detected the coincident signal GW150914 shown in Fig. 1. The initial detection was made by low-latency searches for generic gravitational-wave transients [41] and was reported within three minutes of data acquisition [43]. Subsequently, matched-filter analyses that use relativistic models of compact binary waveforms [44] recovered GW150914 as the most significant event from each detector for the observations reported here. Occurring within the 10-ms intersite

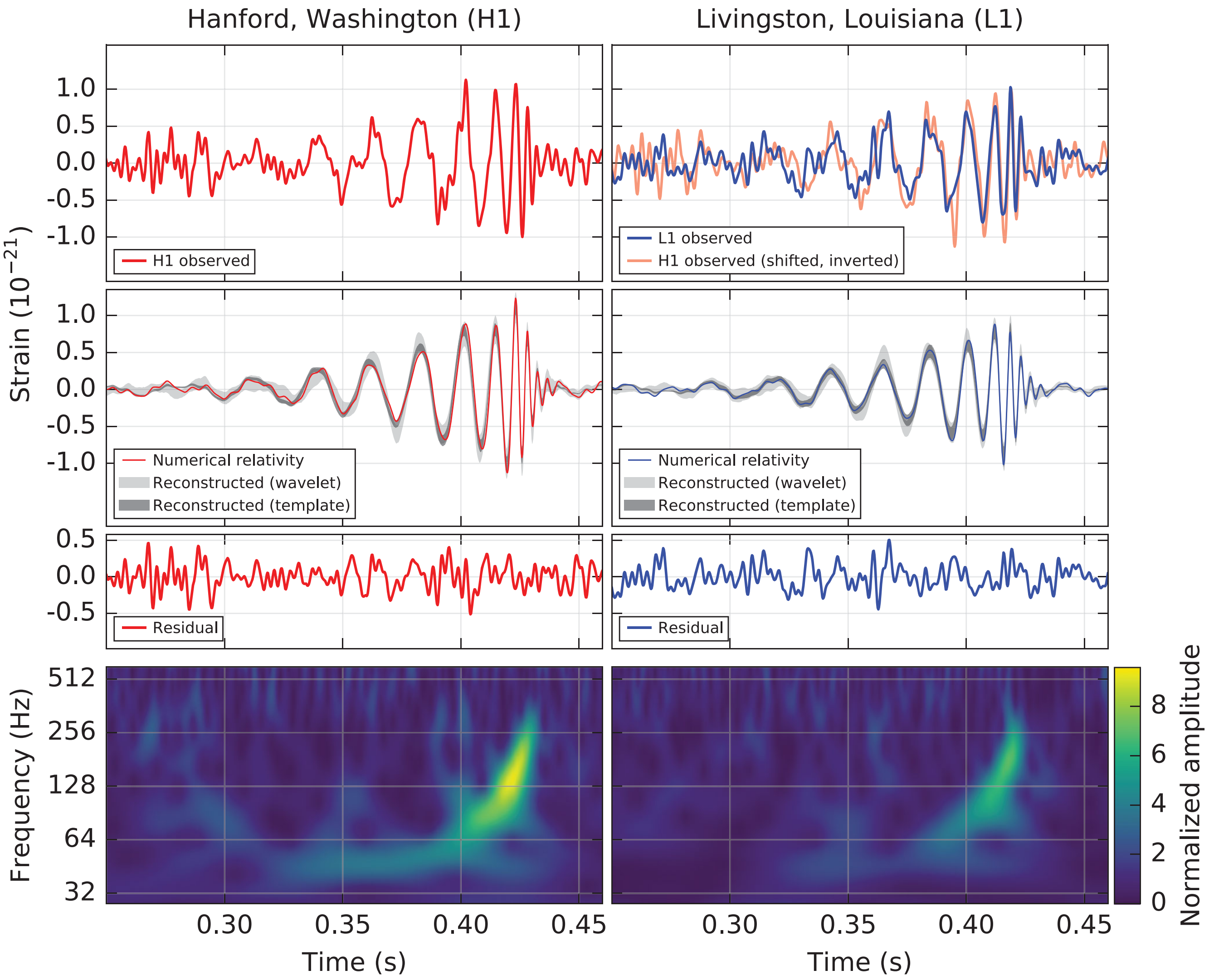

FIG. 1. The gravitational-wave event GW150914 observed by the LIGO Hanford (H1, left column panels) and Livingston (L1, right column panels) detectors. Times are shown relative to September 14, 2015 at 09:50:45 UTC. For visualization, all time series are filtered with a 35–350 Hz bandpass filter to suppress large fluctuations outside the detectors' most sensitive frequency band, and band-reject filters to remove the strong instrumental spectral lines seen in the Fig. 3 spectra. *Top row, left:* H1 strain. *Top row, right:* L1 strain. GW150914 arrived first at L1 and $6.9^{+0.5}_{-0.4}$ ms later at H1; for a visual comparison, the H1 data are also shown, shifted in time by this amount and inverted (to account for the detectors' relative orientations). *Second row:* Gravitational-wave strain projected onto each detector in the 35–350 Hz band. Solid lines show a numerical relativity waveform for a system with parameters consistent with those recovered from GW150914 [37,38] confirmed to 99.9% by an independent calculation based on [15]. Shaded areas show 90% credible regions for two independent waveform reconstructions. One (dark gray) models the signal using binary black hole template waveforms [39]. The other (light gray) does not use an astrophysical model, but instead calculates the strain signal as a linear combination of sine-Gaussian wavelets [40,41]. These reconstructions have a 94% overlap, as shown in [39]. *Third row:* Residuals after subtracting the filtered numerical relativity waveform from the filtered detector time series. *Bottom row:*A time-frequency representation [42] of the strain data, showing the signal frequency increasing over time.

propagation time, the events have a combined signal-to-noise ratio (SNR) of 24 [45].

Only the LIGO detectors were observing at the time of GW150914. The Virgo detector was being upgraded, and GEO 600, though not sufficiently sensitive to detect this event, was operating but not in observational mode. With only two detectors the source position is primarily determined by the relative arrival time and localized to an area of approximately 600 deg$^2$ (90% credible region) [39,46].

The basic features of GW150914 point to it being produced by the coalescence of two black holes—i.e., their orbital inspiral and merger, and subsequent final black hole ringdown. Over 0.2 s, the signal increases in frequency and amplitude in about 8 cycles from 35 to 150 Hz, where the amplitude reaches a maximum. The most plausible explanation for this evolution is the inspiral of two orbiting masses, $m_1$ and $m_2$, due to gravitational-wave emission. At the lower frequencies, such evolution is characterized by the chirp mass [11]

$$\mathcal{M} = \frac{(m_1 m_2)^{3/5}}{(m_1+m_2)^{1/5}} = \frac{c^3}{G}\left[\frac{5}{96}\pi^{-8/3} f^{-11/3} \dot{f}\right]^{3/5},$$

where $f$ and $\dot{f}$ are the observed frequency and its time derivative and $G$ and $c$ are the gravitational constant and speed of light. Estimating $f$ and $\dot{f}$ from the data in Fig. 1, we obtain a chirp mass of $\mathcal{M} \simeq 30 M_\odot$, implying that the total mass $M = m_1 + m_2$ is $\gtrsim 70 M_\odot$ in the detector frame. This bounds the sum of the Schwarzschild radii of the binary components to $2GM/c^2 \gtrsim 210$ km. To reach an orbital frequency of 75 Hz (half the gravitational-wave frequency) the objects must have been very close and very compact; equal Newtonian point masses orbiting at this frequency would be only $\simeq$350 km apart. A pair of neutron stars, while compact, would not have the required mass, while a black hole neutron star binary with the deduced chirp mass would have a very large total mass, and would thus merge at much lower frequency. This leaves black holes as the only known objects compact enough to reach an orbital frequency of 75 Hz without contact. Furthermore, the decay of the waveform after it peaks is consistent with the damped oscillations of a black hole relaxing to a final stationary Kerr configuration. Below, we present a general-relativistic analysis of GW150914; Fig. 2 shows the calculated waveform using the resulting source parameters.

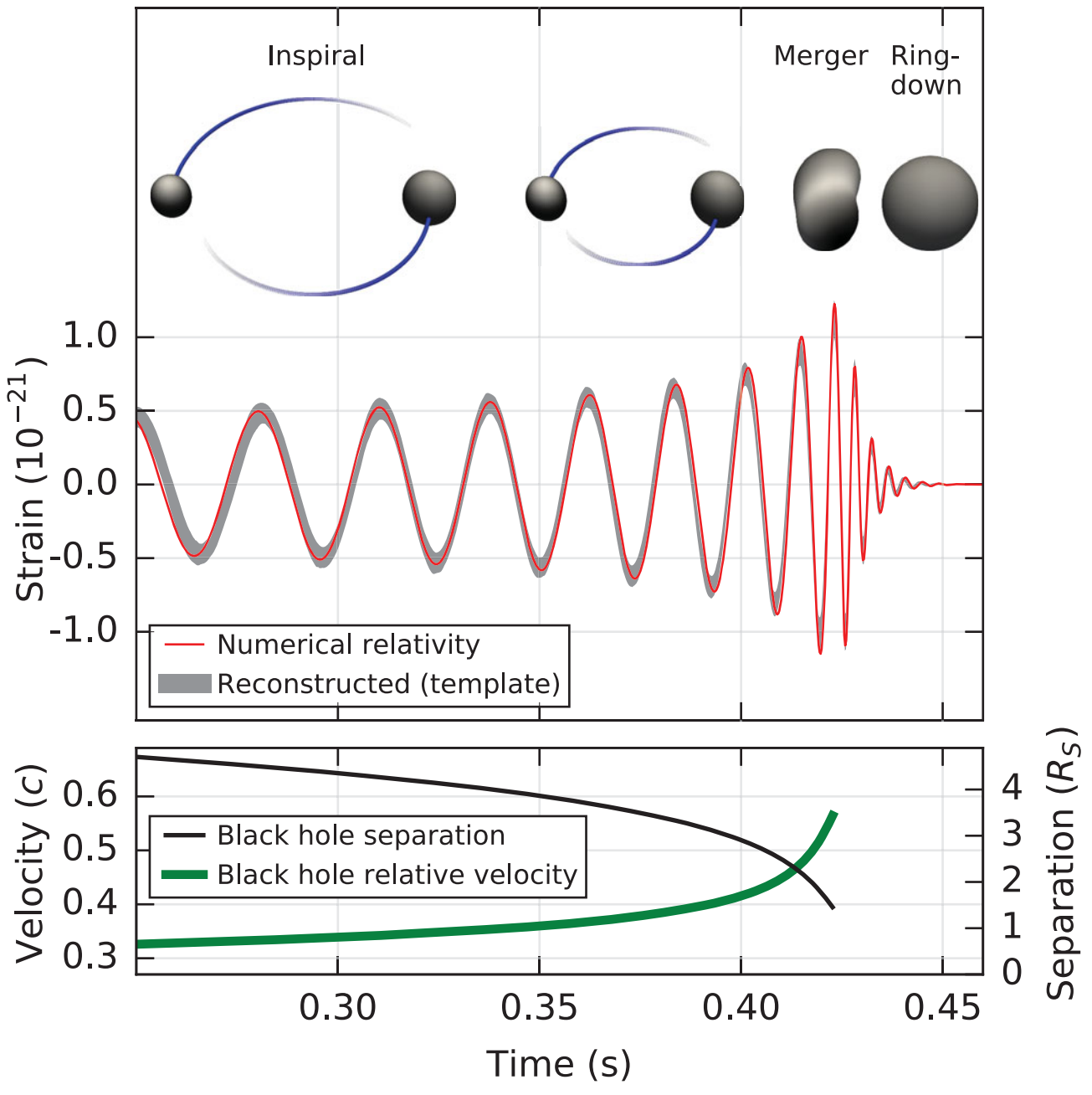

FIG. 2. *Top:* Estimated gravitational-wave strain amplitude from GW150914 projected onto H1. This shows the full bandwidth of the waveforms, without the filtering used for Fig. 1. The inset images show numerical relativity models of the black hole horizons as the black holes coalesce. *Bottom:* The Keplerian effective black hole separation in units of Schwarzschild radii ($R_S = 2GM/c^2$) and the effective relative velocity given by the post-Newtonian parameter $v/c = (GM\pi f/c^3)^{1/3}$, where $f$ is the gravitational-wave frequency calculated with numerical relativity and $M$ is the total mass (value from Table I).

## III. DETECTORS

Gravitational-wave astronomy exploits multiple, widely separated detectors to distinguish gravitational waves from local instrumental and environmental noise, to provide source sky localization, and to measure wave polarizations. The LIGO sites each operate a single Advanced LIGO detector [33], a modified Michelson interferometer (see Fig. 3) that measures gravitational-wave strain as a difference in length of its orthogonal arms. Each arm is formed by two mirrors, acting as test masses, separated by $L_x = L_y = L = 4$ km. A passing gravitational wave effectively alters the arm lengths such that the measured difference is $\Delta L(t) = \delta L_x - \delta L_y = h(t)L$, where $h$ is the gravitational-wave strain amplitude projected onto the detector. This differential length variation alters the phase difference between the two light fields returning to the beam splitter, transmitting an optical signal proportional to the gravitational-wave strain to the output photodetector.

To achieve sufficient sensitivity to measure gravitational waves, the detectors include several enhancements to the basic Michelson interferometer. First, each arm contains a resonant optical cavity, formed by its two test mass mirrors, that multiplies the effect of a gravitational wave on the light phase by a factor of 300 [48]. Second, a partially transmissive power-recycling mirror at the input provides additional resonant buildup of the laser light in the interferometer as a whole [49,50]: 20 W of laser input is increased to 700 W incident on the beam splitter, which is further increased to 100 kW circulating in each arm cavity. Third, a partially transmissive signal-recycling mirror at the output optimizes the gravitational-wave signal extraction by broadening the bandwidth of the arm cavities [51,52]. The interferometer is illuminated with a 1064-nm wavelength Nd:YAG laser, stabilized in amplitude, frequency, and beam geometry [53,54]. The gravitational-wave signal is extracted at the output port using a homodyne readout [55].

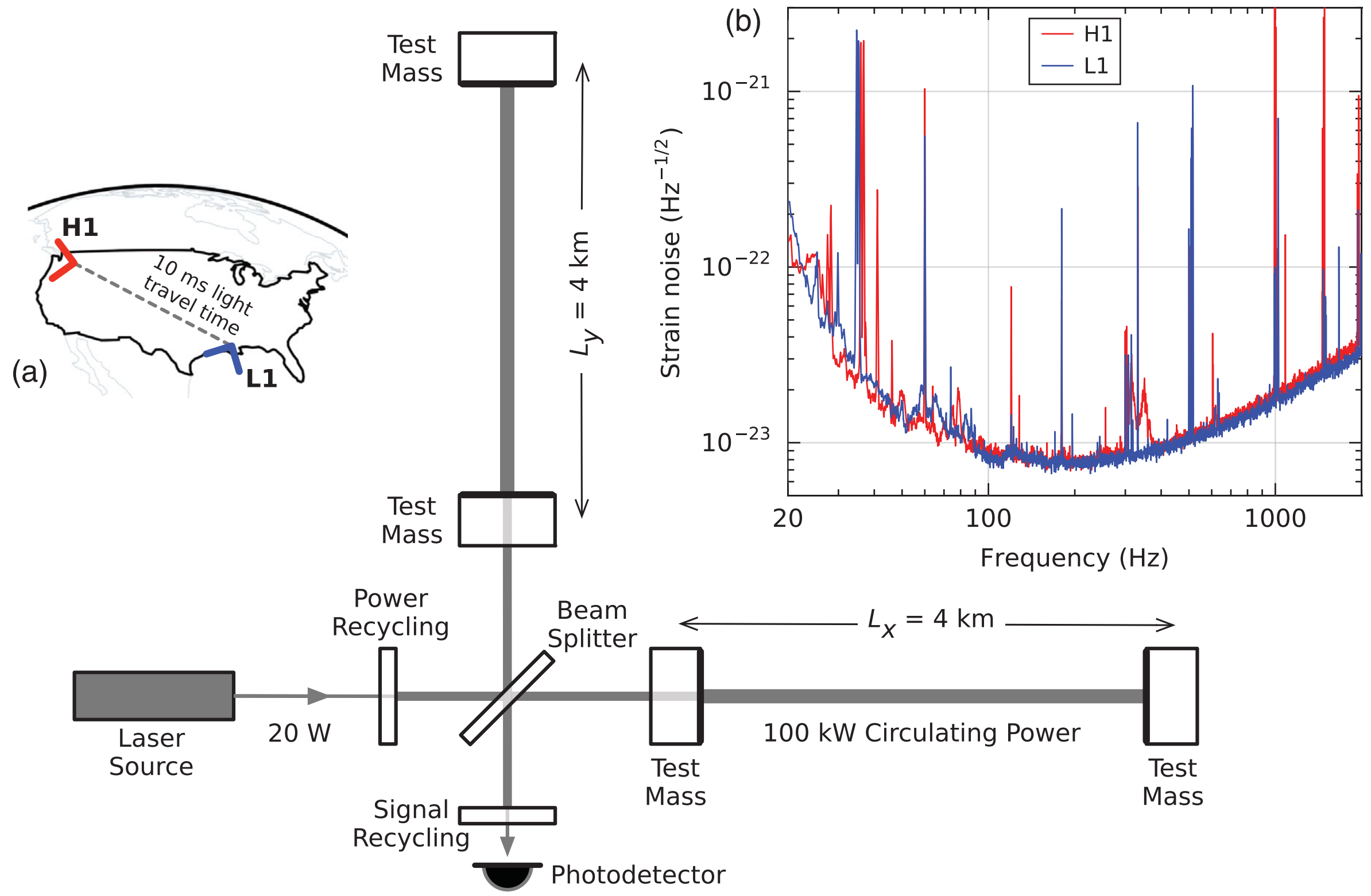

FIG. 3. Simplified diagram of an Advanced LIGO detector (not to scale). A gravitational wave propagating orthogonally to the detector plane and linearly polarized parallel to the 4-km optical cavities will have the effect of lengthening one 4-km arm and shortening the other during one half-cycle of the wave; these length changes are reversed during the other half-cycle. The output photodetector records these differential cavity length variations. While a detector's directional response is maximal for this case, it is still significant for most other angles of incidence or polarizations (gravitational waves propagate freely through the Earth). *Inset (a):* Location and orientation of the LIGO detectors at Hanford, WA (H1) and Livingston, LA (L1). *Inset (b):* The instrument noise for each detector near the time of the signal detection; this is an amplitude spectral density, expressed in terms of equivalent gravitational-wave strain amplitude. The sensitivity is limited by photon shot noise at frequencies above 150 Hz, and by a superposition of other noise sources at lower frequencies [47]. Narrow-band features include calibration lines (33–38, 330, and 1080 Hz), vibrational modes of suspension fibers (500 Hz and harmonics), and 60 Hz electric power grid harmonics.

These interferometry techniques are designed to maximize the conversion of strain to optical signal, thereby minimizing the impact of photon shot noise (the principal noise at high frequencies). High strain sensitivity also requires that the test masses have low displacement noise, which is achieved by isolating them from seismic noise (low frequencies) and designing them to have low thermal noise (intermediate frequencies). Each test mass is suspended as the final stage of a quadruple-pendulum system [56], supported by an active seismic isolation platform [57]. These systems collectively provide more than 10 orders of magnitude of isolation from ground motion for frequencies above 10 Hz. Thermal noise is minimized by using low-mechanical-loss materials in the test masses and their suspensions: the test masses are 40-kg fused silica substrates with low-loss dielectric optical coatings [58,59], and are suspended with fused silica fibers from the stage above [60].

To minimize additional noise sources, all components other than the laser source are mounted on vibration isolation stages in ultrahigh vacuum. To reduce optical phase fluctuations caused by Rayleigh scattering, the pressure in the 1.2-m diameter tubes containing the arm-cavity beams is maintained below 1 $\mu$Pa.

Servo controls are used to hold the arm cavities on resonance [61] and maintain proper alignment of the optical components [62]. The detector output is calibrated in strain by measuring its response to test mass motion induced by photon pressure from a modulated calibration laser beam [63]. The calibration is established to an uncertainty ($1\sigma$) of less than 10% in amplitude and 10 degrees in phase, and is continuously monitored with calibration laser excitations at selected frequencies. Two alternative methods are used to validate the absolute calibration, one referenced to the main laser wavelength and the other to a radio-frequency oscillator [64]. Additionally, the detector response to gravitational waves is tested by injecting simulated waveforms with the calibration laser.

To monitor environmental disturbances and their influence on the detectors, each observatory site is equipped with an array of sensors: seismometers, accelerometers, microphones, magnetometers, radio receivers, weather sensors, ac-power line monitors, and a cosmic-ray detector [65]. Another $\sim 10^5$ channels record the interferometer's operating point and the state of the control systems. Data collection is synchronized to Global Positioning System (GPS) time to better than 10 $\mu$s [66]. Timing accuracy is verified with an atomic clock and a secondary GPS receiver at each observatory site.

In their most sensitive band, 100–300 Hz, the current LIGO detectors are 3 to 5 times more sensitive to strain than initial LIGO [67]; at lower frequencies, the improvement is even greater, with more than ten times better sensitivity below 60 Hz. Because the detectors respond proportionally to gravitational-wave amplitude, at low redshift the volume of space to which they are sensitive increases as the cube of strain sensitivity. For binary black holes with masses similar to GW150914, the space-time volume surveyed by the observations reported here surpasses previous observations by an order of magnitude [68].

## IV. DETECTOR VALIDATION

Both detectors were in steady state operation for several hours around GW150914. All performance measures, in particular their average sensitivity and transient noise behavior, were typical of the full analysis period [69,70].

Exhaustive investigations of instrumental and environmental disturbances were performed, giving no evidence to suggest that GW150914 could be an instrumental artifact [69]. The detectors' susceptibility to environmental disturbances was quantified by measuring their response to specially generated magnetic, radio-frequency, acoustic, and vibration excitations. These tests indicated that any external disturbance large enough to have caused the observed signal would have been clearly recorded by the array of environmental sensors. None of the environmental sensors recorded any disturbances that evolved in time and frequency like GW150914, and all environmental fluctuations during the second that contained GW150914 were too small to account for more than 6% of its strain amplitude. Special care was taken to search for long-range correlated disturbances that might produce nearly simultaneous signals at the two sites. No significant disturbances were found.

The detector strain data exhibit non-Gaussian noise transients that arise from a variety of instrumental mechanisms. Many have distinct signatures, visible in auxiliary data channels that are not sensitive to gravitational waves; such instrumental transients are removed from our analyses [69]. Any instrumental transients that remain in the data are accounted for in the estimated detector backgrounds described below. There is no evidence for instrumental transients that are temporally correlated between the two detectors.

## V. SEARCHES

We present the analysis of 16 days of coincident observations between the two LIGO detectors from September 12 to October 20, 2015. This is a subset of the data from Advanced LIGO's first observational period that ended on January 12, 2016.

GW150914 is confidently detected by two different types of searches. One aims to recover signals from the coalescence of compact objects, using optimal matched filtering with waveforms predicted by general relativity. The other search targets a broad range of generic transient signals, with minimal assumptions about waveforms. These searches use independent methods, and their response to detector noise consists of different, uncorrelated, events. However, strong signals from binary black hole mergers are expected to be detected by both searches.

Each search identifies candidate events that are detected at both observatories consistent with the intersite propagation time. Events are assigned a detection-statistic value that ranks their likelihood of being a gravitational-wave signal. The significance of a candidate event is determined by the search background—the rate at which detector noise produces events with a detection-statistic value equal to or higher than the candidate event. Estimating this background is challenging for two reasons: the detector noise is nonstationary and non-Gaussian, so its properties must be empirically determined; and it is not possible to shield the detector from gravitational waves to directly measure a signal-free background. The specific procedure used to estimate the background is slightly different for the two searches, but both use a time-shift technique: the time stamps of one detector's data are artificially shifted by an offset that is large compared to the intersite propagation time, and a new set of events is produced based on this time-shifted data set. For instrumental noise that is uncorrelated between detectors this is an effective way to estimate the background. In this process a gravitational-wave signal in one detector may coincide with time-shifted noise transients in the other detector, thereby contributing to the background estimate. This leads to an overestimate of the noise background and therefore to a more conservative assessment of the significance of candidate events.

The characteristics of non-Gaussian noise vary between different time-frequency regions. This means that the search backgrounds are not uniform across the space of signals being searched. To maximize sensitivity and provide a better estimate of event significance, the searches sort both their background estimates and their event candidates into different classes according to their time-frequency morphology. The significance of a candidate event is measured against the background of its class. To account for having searched multiple classes, this significance is decreased by a trials factor equal to the number of classes [71].

### A. Generic transient search

Designed to operate without a specific waveform model, this search identifies coincident excess power in time-frequency representations of the detector strain data [43,72], for signal frequencies up to 1 kHz and durations up to a few seconds.

The search reconstructs signal waveforms consistent with a common gravitational-wave signal in both detectors using a multidetector maximum likelihood method. Each event is ranked according to the detection statistic $\eta_c = \sqrt{2E_c/(1 + E_n/E_c)}$, where $E_c$ is the dimensionless coherent signal energy obtained by cross-correlating the two reconstructed waveforms, and $E_n$ is the dimensionless residual noise energy after the reconstructed signal is subtracted from the data. The statistic $\eta_c$ thus quantifies the SNR of the event and the consistency of the data between the two detectors.

Based on their time-frequency morphology, the events are divided into three mutually exclusive search classes, as described in [41]: events with time-frequency morphology of known populations of noise transients (class C1), events with frequency that increases with time (class C3), and all remaining events (class C2).

Detected with $\eta_c = 20.0$, GW150914 is the strongest event of the entire search. Consistent with its coalescence signal signature, it is found in the search class C3 of events with increasing time-frequency evolution. Measured on a background equivalent to over 67 400 years of data and including a trials factor of 3 to account for the search classes, its false alarm rate is lower than 1 in 22 500 years. This corresponds to a probability $< 2 \times 10^{-6}$ of observing one or more noise events as strong as GW150914 during the analysis time, equivalent to $4.6\sigma$. The left panel of Fig. 4 shows the C3 class results and background.

The selection criteria that define the search class C3 reduce the background by introducing a constraint on the signal morphology. In order to illustrate the significance of GW150914 against a background of events with arbitrary shapes, we also show the results of a search that uses the same set of events as the one described above but without this constraint. Specifically, we use only two search classes: the C1 class and the union of C2 and C3 classes (C2 + C3). In this two-class search the GW150914 event is found in the C2 + C3 class. The left panel of Fig. 4 shows the C2 + C3 class results and background. In the background of this class there are four events with $\eta_c \geq 32.1$, yielding a false alarm rate for GW150914 of 1 in 8 400 years. This corresponds to a false alarm probability of $5 \times 10^{-6}$ equivalent to $4.4\sigma$.

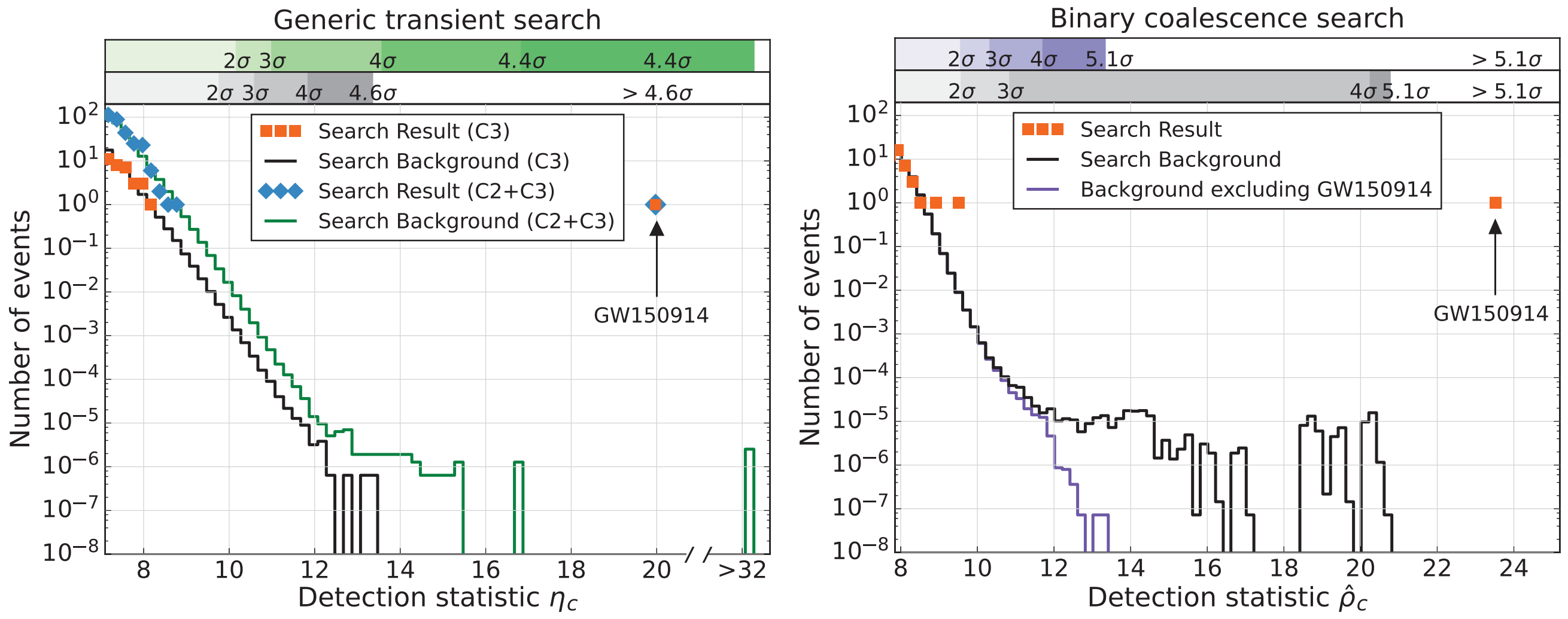

FIG. 4. Search results from the generic transient search (left) and the binary coalescence search (right). These histograms show the number of candidate events (orange markers) and the mean number of background events (black lines) in the search class where GW150914 was found as a function of the search detection statistic and with a bin width of 0.2. The scales on the top give the significance of an event in Gaussian standard deviations based on the corresponding noise background. The significance of GW150914 is greater than $5.1\sigma$ and $4.6\sigma$ for the binary coalescence and the generic transient searches, respectively. *Left:* Along with the primary search (C3) we also show the results (blue markers) and background (green curve) for an alternative search that treats events independently of their frequency evolution (C2 + C3). The classes C2 and C3 are defined in the text. *Right:* The tail in the black-line background of the binary coalescence search is due to random coincidences of GW150914 in one detector with noise in the other detector. (This type of event is practically absent in the generic transient search background because they do not pass the time-frequency consistency requirements used in that search.) The purple curve is the background excluding those coincidences, which is used to assess the significance of the second strongest event.

For robustness and validation, we also use other generic transient search algorithms [41]. A different search [73] and a parameter estimation follow-up [74] detected GW150914 with consistent significance and signal parameters.

### B. Binary coalescence search

This search targets gravitational-wave emission from binary systems with individual masses from 1 to $99M_\odot$, total mass less than $100M_\odot$, and dimensionless spins up to 0.99 [44]. To model systems with total mass larger than $4M_\odot$, we use the effective-one-body formalism [75], which combines results from the post-Newtonian approach [11,76] with results from black hole perturbation theory and numerical relativity. The waveform model [77,78] assumes that the spins of the merging objects are aligned with the orbital angular momentum, but the resulting templates can, nonetheless, effectively recover systems with misaligned spins in the parameter region of GW150914 [44]. Approximately 250 000 template waveforms are used to cover this parameter space.

The search calculates the matched-filter signal-to-noise ratio $\rho(t)$ for each template in each detector and identifies maxima of $\rho(t)$ with respect to the time of arrival of the signal [79–81]. For each maximum we calculate a chi-squared statistic $\chi_r^2$ to test whether the data in several different frequency bands are consistent with the matching template [82]. Values of $\chi_r^2$ near unity indicate that the signal is consistent with a coalescence. If $\chi_r^2$ is greater than unity, $\rho(t)$ is reweighted as $\hat{\rho} = \rho/\{[1 + (\chi_r^2)^3]/2\}^{1/6}$ [83,84]. The final step enforces coincidence between detectors by selecting event pairs that occur within a 15-ms window and come from the same template. The 15-ms window is determined by the 10-ms intersite propagation time plus 5 ms for uncertainty in arrival time of weak signals. We rank coincident events based on the quadrature sum $\hat{\rho}_c$ of the $\hat{\rho}$ from both detectors [45].

To produce background data for this search the SNR maxima of one detector are time shifted and a new set of coincident events is computed. Repeating this procedure $\sim 10^7$ times produces a noise background analysis time equivalent to 608 000 years.

To account for the search background noise varying across the target signal space, candidate and background events are divided into three search classes based on template length. The right panel of Fig. 4 shows the background for the search class of GW150914. The GW150914 detection-statistic value of $\hat{\rho}_c = 23.6$ is larger than any background event, so only an upper bound can be placed on its false alarm rate. Across the three search classes this bound is 1 in 203 000 years. This translates to a false alarm probability $< 2 \times 10^{-7}$, corresponding to $5.1\sigma$.

A second, independent matched-filter analysis that uses a different method for estimating the significance of its events [85,86], also detected GW150914 with identical signal parameters and consistent significance.

TABLE I. Source parameters for GW150914. We report median values with 90% credible intervals that include statistical errors, and systematic errors from averaging the results of different waveform models. Masses are given in the source frame; to convert to the detector frame multiply by $(1 + z)$ [90]. The source redshift assumes standard cosmology [91].

| | |
|---|---|
| Primary black hole mass | $36^{+5}_{-4} M_\odot$ |
| Secondary black hole mass | $29^{+4}_{-4} M_\odot$ |
| Final black hole mass | $62^{+4}_{-4} M_\odot$ |
| Final black hole spin | $0.67^{+0.05}_{-0.07}$ |
| Luminosity distance | $410^{+160}_{-180}$ Mpc |
| Source redshift $z$ | $0.09^{+0.03}_{-0.04}$ |

When an event is confidently identified as a real gravitational-wave signal, as for GW150914, the background used to determine the significance of other events is reestimated without the contribution of this event. This is the background distribution shown as a purple line in the right panel of Fig. 4. Based on this, the second most significant event has a false alarm rate of 1 per 2.3 years and corresponding Poissonian false alarm probability of 0.02. Waveform analysis of this event indicates that if it is astrophysical in origin it is also a binary black hole merger [44].

## VI. SOURCE DISCUSSION

The matched-filter search is optimized for detecting signals, but it provides only approximate estimates of the source parameters. To refine them we use general relativity-based models [77,78,87,88], some of which include spin precession, and for each model perform a coherent Bayesian analysis to derive posterior distributions of the source parameters [89]. The initial and final masses, final spin, distance, and redshift of the source are shown in Table I. The spin of the primary black hole is constrained to be $< 0.7$ (90% credible interval) indicating it is not maximally spinning, while the spin of the secondary is only weakly constrained. These source parameters are discussed in detail in [39]. The parameter uncertainties include statistical errors and systematic errors from averaging the results of different waveform models.

Using the fits to numerical simulations of binary black hole mergers in [92,93], we provide estimates of the mass and spin of the final black hole, the total energy radiated in gravitational waves, and the peak gravitational-wave luminosity [39]. The estimated total energy radiated in gravitational waves is $3.0^{+0.5}_{-0.5} M_\odot c^2$. The system reached a peak gravitational-wave luminosity of $3.6^{+0.5}_{-0.4} \times 10^{56}$ erg/s, equivalent to $200^{+30}_{-20} M_\odot c^2/\mathrm{s}$.

Several analyses have been performed to determine whether or not GW150914 is consistent with a binary black hole system in general relativity [94]. A first consistency check involves the mass and spin of the final black hole. In general relativity, the end product of a black hole binary coalescence is a Kerr black hole, which is fully described by its mass and spin. For quasicircular inspirals, these are predicted uniquely by Einstein's equations as a function of the masses and spins of the two progenitor black holes. Using fitting formulas calibrated to numerical relativity simulations [92], we verified that the remnant mass and spin deduced from the early stage of the coalescence and those inferred independently from the late stage are consistent with each other, with no evidence for disagreement from general relativity.

Within the post-Newtonian formalism, the phase of the gravitational waveform during the inspiral can be expressed as a power series in $f^{1/3}$. The coefficients of this expansion can be computed in general relativity. Thus, we can test for consistency with general relativity [95,96] by allowing the coefficients to deviate from the nominal values, and seeing if the resulting waveform is consistent with the data. In this second check [94] we place constraints on these deviations, finding no evidence for violations of general relativity.

Finally, assuming a modified dispersion relation for gravitational waves [97], our observations constrain the Compton wavelength of the graviton to be $\lambda_g > 10^{13}$ km, which could be interpreted as a bound on the graviton mass $m_g < 1.2 \times 10^{-22}$ eV$/c^2$. This improves on Solar System and binary pulsar bounds [98,99] by factors of a few and a thousand, respectively, but does not improve on the model-dependent bounds derived from the dynamics of Galaxy clusters [100] and weak lensing observations [101]. In summary, all three tests are consistent with the predictions of general relativity in the strong-field regime of gravity.

GW150914 demonstrates the existence of stellar-mass black holes more massive than $\simeq 25 M_\odot$, and establishes that binary black holes can form in nature and merge within a Hubble time. Binary black holes have been predicted to form both in isolated binaries [102–104] and in dense environments by dynamical interactions [105–107]. The formation of such massive black holes from stellar evolution requires weak massive-star winds, which are possible in stellar environments with metallicity lower than $\simeq 1/2$ the solar value [108,109]. Further astrophysical implications of this binary black hole discovery are discussed in [110].

These observational results constrain the rate of stellar-mass binary black hole mergers in the local universe. Using several different models of the underlying binary black hole mass distribution, we obtain rate estimates ranging from 2–400 $\text{Gpc}^{-3}\ \text{yr}^{-1}$ in the comoving frame [111–113]. This is consistent with a broad range of rate predictions as reviewed in [114], with only the lowest event rates being excluded.

Binary black hole systems at larger distances contribute to a stochastic background of gravitational waves from the superposition of unresolved systems. Predictions for such a background are presented in [115]. If the signal from such a population were detected, it would provide information about the evolution of such binary systems over the history of the universe.

## VII. OUTLOOK

Further details about these results and associated data releases are available at [116]. Analysis results for the entire first observational period will be reported in future publications. Efforts are under way to enhance significantly the global gravitational-wave detector network [117]. These include further commissioning of the Advanced LIGO detectors to reach design sensitivity, which will allow detection of binaries like GW150914 with 3 times higher SNR. Additionally, Advanced Virgo, KAGRA, and a possible third LIGO detector in India [118] will extend the network and significantly improve the position reconstruction and parameter estimation of sources.

## VIII. CONCLUSION

The LIGO detectors have observed gravitational waves from the merger of two stellar-mass black holes. The detected waveform matches the predictions of general relativity for the inspiral and merger of a pair of black holes and the ringdown of the resulting single black hole. These observations demonstrate the existence of binary stellar-mass black hole systems. This is the first direct detection of gravitational waves and the first observation of a binary black hole merger.

## ACKNOWLEDGMENTS

The authors gratefully acknowledge the support of the United States National Science Foundation (NSF) for the construction and operation of the LIGO Laboratory and Advanced LIGO as well as the Science and Technology Facilities Council (STFC) of the United Kingdom, the Max-Planck Society (MPS), and the State of Niedersachsen, Germany, for support of the construction of Advanced LIGO and construction and operation of the GEO 600 detector. Additional support for Advanced LIGO was provided by the Australian Research Council. The authors gratefully acknowledge the Italian Istituto Nazionale di Fisica Nucleare (INFN), the French Centre National de la Recherche Scientifique (CNRS), and the Foundation for Fundamental Research on Matter supported by the Netherlands Organisation for Scientific Research, for the construction and operation of the Virgo detector, and for the creation and support of the EGO consortium. The authors also gratefully acknowledge research support from these agencies as well as by the Council of Scientific and Industrial Research of India, Department of Science and Technology, India, Science & Engineering Research Board (SERB), India, Ministry of Human Resource Development, India, the Spanish Ministerio de Economía y Competitividad, the Conselleria d'Economia i Competitivitat and Conselleria d'Educació, Cultura i

Universitats of the Govern de les Illes Balears, the National Science Centre of Poland, the European Commission, the Royal Society, the Scottish Funding Council, the Scottish Universities Physics Alliance, the Hungarian Scientific Research Fund (OTKA), the Lyon Institute of Origins (LIO), the National Research Foundation of Korea, Industry Canada and the Province of Ontario through the Ministry of Economic Development and Innovation, the Natural Sciences and Engineering Research Council of Canada, Canadian Institute for Advanced Research, the Brazilian Ministry of Science, Technology, and Innovation, Russian Foundation for Basic Research, the Leverhulme Trust, the Research Corporation, Ministry of Science and Technology (MOST), Taiwan, and the Kavli Foundation. The authors gratefully acknowledge the support of the NSF, STFC, MPS, INFN, CNRS and the State of Niedersachsen, Germany, for provision of computational resources. This article has been assigned the document numbers LIGO-P150914 and VIR-0015A-16.

B. P. Abbott,[1] R. Abbott,[1] T. D. Abbott,[2] M. R. Abernathy,[1] F. Acernese,[3,4] K. Ackley,[5] C. Adams,[6] T. Adams,[7] P. Addesso,[3] R. X. Adhikari,[1] V. B. Adya,[8] C. Affeldt,[8] M. Agathos,[9] K. Agatsuma,[9] N. Aggarwal,[10] O. D. Aguiar,[11] L. Aiello,[12,13] A. Ain,[14] P. Ajith,[15] B. Allen,[8,16,17] A. Allocca,[18,19] P. A. Altin,[20] S. B. Anderson,[1] W. G. Anderson,[16] K. Arai,[1] M. A. Arain,[5] M. C. Araya,[1] C. C. Arceneaux,[21] J. S. Areeda,[22] N. Arnaud,[23] K. G. Arun,[24] S. Ascenzi,[25,13] G. Ashton,[26] M. Ast,[27] S. M. Aston,[6] P. Astone,[28] P. Aufmuth,[8] C. Aulbert,[8] S. Babak,[29] P. Bacon,[30] M. K. M. Bader,[9] P. T. Baker,[31] F. Baldaccini,[32,33] G. Ballardin,[34] S. W. Ballmer,[35] J. C. Barayoga,[1] S. E. Barclay,[36] B. C. Barish,[1] D. Barker,[37] F. Barone,[3,4] B. Barr,[36] L. Barsotti,[10] M. Barsuglia,[30] D. Barta,[38] J. Bartlett,[37] M. A. Barton,[37] I. Bartos,[39] R. Bassiri,[40] A. Basti,[18,19] J. C. Batch,[37] C. Baune,[8] V. Bavigadda,[34] M. Bazzan,[41,42] B. Behnke,[29] M. Bejger,[43] C. Belczynski,[44] A. S. Bell,[36] C. J. Bell,[36] B. K. Berger,[1] J. Bergman,[37] G. Bergmann,[8] C. P. L. Berry,[45] D. Bersanetti,[46,47] A. Bertolini,[9] J. Betzwieser,[6] S. Bhagwat,[35] R. Bhandare,[48] I. A. Bilenko,[49] G. Billingsley,[1] J. Birch,[6] R. Birney,[50] O. Birnholtz,[8] S. Biscans,[10] A. Bisht,[8,17] M. Bitossi,[34] C. Biwer,[35] M. A. Bizouard,[23] J. K. Blackburn,[1] C. D. Blair,[51] D. G. Blair,[51] R. M. Blair,[37] S. Bloemen,[52] O. Bock,[8] T. P. Bodiya,[10] M. Boer,[53] G. Bogaert,[53] C. Bogan,[8] A. Bohe,[29] P. Bojtos,[54] C. Bond,[45] F. Bondu,[55] R. Bonnand,[7] B. A. Boom,[9] R. Bork,[1] V. Boschi,[18,19] S. Bose,[56,14] Y. Bouffanais,[30] A. Bozzi,[34] C. Bradaschia,[19] P. R. Brady,[16] V. B. Braginsky,[49] M. Branchesi,[57,58] J. E. Brau,[59] T. Briant,[60] A. Brillet,[53] M. Brinkmann,[8] V. Brisson,[23] P. Brockill,[16] A. F. Brooks,[1] D. A. Brown,[35] D. D. Brown,[45] N. M. Brown,[10] C. C. Buchanan,[2] A. Buikema,[10] T. Bulik,[44] H. J. Bulten,[61,9] A. Buonanno,[29,62] D. Buskulic,[7] C. Buy,[30] R. L. Byer,[40] M. Cabero,[8] L. Cadonati,[63] G. Cagnoli,[64,65] C. Cahillane,[1] J. Calderón Bustillo,[66,63] T. Callister,[1] E. Calloni,[67,4] J. B. Camp,[68] K. C. Cannon,[69] J. Cao,[70] C. D. Capano,[8] E. Capocasa,[30] F. Carbognani,[34] S. Caride,[71] J. Casanueva Diaz,[23] C. Casentini,[25,13] S. Caudill,[16] M. Cavaglià,[21] F. Cavalier,[23] R. Cavalieri,[34] G. Cella,[19] C. B. Cepeda,[1] L. Cerboni Baiardi,[57,58] G. Cerretani,[18,19] E. Cesarini,[25,13] R. Chakraborty,[1] T. Chalermsongsak,[1] S. J. Chamberlin,[72] M. Chan,[36] S. Chao,[73] P. Charlton,[74] E. Chassande-Mottin,[30] H. Y. Chen,[75] Y. Chen,[76] C. Cheng,[73] A. Chincarini,[47] A. Chiummo,[34] H. S. Cho,[77] M. Cho,[62] J. H. Chow,[20] N. Christensen,[78] Q. Chu,[51] S. Chua,[60] S. Chung,[51] G. Ciani,[5] F. Clara,[37] J. A. Clark,[63] F. Cleva,[53] E. Coccia,[25,12,13] P.-F. Cohadon,[60] A. Colla,[79,28] C. G. Collette,[80] L. Cominsky,[81] M. Constancio Jr.,[11] A. Conte,[79,28] L. Conti,[42] D. Cook,[37] T. R. Corbitt,[2] N. Cornish,[31] A. Corsi,[71] S. Cortese,[34] C. A. Costa,[11] M. W. Coughlin,[78] S. B. Coughlin,[82] J.-P. Coulon,[53] S. T. Countryman,[39] P. Couvares,[1] E. E. Cowan,[63] D. M. Coward,[51] M. J. Cowart,[6] D. C. Coyne,[1] R. Coyne,[71] K. Craig,[36] J. D. E. Creighton,[16] T. D. Creighton,[83] J. Cripe,[2] S. G. Crowder,[84] A. M. Cruise,[45] A. Cumming,[36] L. Cunningham,[36] E. Cuoco,[34] T. Dal Canton,[8] S. L. Danilishin,[36] S. D'Antonio,[13] K. Danzmann,[17,8] N. S. Darman,[85] C. F. Da Silva Costa,[5] V. Dattilo,[34] I. Dave,[48] H. P. Daveloza,[83] M. Davier,[23] G. S. Davies,[36] E. J. Daw,[86] R. Day,[34] S. De,[35] D. DeBra,[40] G. Debreczeni,[38] J. Degallaix,[65] M. De Laurentis,[67,4] S. Deléglise,[60] W. Del Pozzo,[45] T. Denker,[8,17] T. Dent,[8] H. Dereli,[53] V. Dergachev,[1] R. T. DeRosa,[6] R. De Rosa,[67,4] R. DeSalvo,[87] S. Dhurandhar,[14] M. C. Díaz,[83] L. Di Fiore,[4] M. Di Giovanni,[79,28] A. Di Lieto,[18,19] S. Di Pace,[79,28] I. Di Palma,[29,8] A. Di Virgilio,[19] G. Dojcinoski,[88] V. Dolique,[65] F. Donovan,[10] K. L. Dooley,[21] S. Doravari,[6,8] R. Douglas,[36] T. P. Downes,[16] M. Drago,[8,89,90] R. W. P. Drever,[1] J. C. Driggers,[37] Z. Du,[70] M. Ducrot,[7] S. E. Dwyer,[37] T. B. Edo,[86] M. C. Edwards,[78] A. Effler,[6] H.-B. Eggenstein,[8] P. Ehrens,[1] J. Eichholz,[5] S. S. Eikenberry,[5] W. Engels,[76] R. C. Essick,[10] T. Etzel,[1] M. Evans,[10] T. M. Evans,[6] R. Everett,[72] M. Factourovich,[39] V. Fafone,[25,13,12] H. Fair,[35] S. Fairhurst,[91] X. Fan,[70] Q. Fang,[51] S. Farinon,[47] B. Farr,[75] W. M. Farr,[45] M. Favata,[88] M. Fays,[91] H. Fehrmann,[8] M. M. Fejer,[40] D. Feldbaum,[5] I. Ferrante,[18,19] E. C. Ferreira,[11] F. Ferrini,[34] F. Fidecaro,[18,19] L. S. Finn,[72] I. Fiori,[34] D. Fiorucci,[30] R. P. Fisher,[35] R. Flaminio,[65,92] M. Fletcher,[36] H. Fong,[69] J.-D. Fournier,[53] S. Franco,[23] S. Frasca,[79,28] F. Frasconi,[19] M. Frede,[8] Z. Frei,[54] A. Freise,[45] R. Frey,[59] V. Frey,[23] T. T. Fricke,[8] P. Fritschel,[10] V. V. Frolov,[6] P. Fulda,[5] M. Fyffe,[6] H. A. G. Gabbard,[21] J. R. Gair,[93] L. Gammaitoni,[32,33] S. G. Gaonkar,[14] F. Garufi,[67,4] A. Gatto,[30] G. Gaur,[94,95]

N. Gehrels,[68] G. Gemme,[47] B. Gendre,[53] E. Genin,[34] A. Gennai,[19] J. George,[48] L. Gergely,[96] V. Germain,[7] Abhirup Ghosh,[15] Archisman Ghosh,[15] S. Ghosh,[52,9] J. A. Giaime,[2,6] K. D. Giardina,[6] A. Giazotto,[19] K. Gill,[97] A. Glaefke,[36] J. R. Gleason,[5] E. Goetz,[98] R. Goetz,[5] L. Gondan,[54] G. González,[2] J. M. Gonzalez Castro,[18,19] A. Gopakumar,[99] N. A. Gordon,[36] M. L. Gorodetsky,[49] S. E. Gossan,[1] M. Gosselin,[34] R. Gouaty,[7] C. Graef,[36] P. B. Graff,[62] M. Granata,[65] A. Grant,[36] S. Gras,[10] C. Gray,[37] G. Greco,[57,58] A. C. Green,[45] R. J. S. Greenhalgh,[100] P. Groot,[52] H. Grote,[8] S. Grunewald,[29] G. M. Guidi,[57,58] X. Guo,[70] A. Gupta,[14] M. K. Gupta,[95] K. E. Gushwa,[1] E. K. Gustafson,[1] R. Gustafson,[98] J. J. Hacker,[22] B. R. Hall,[56] E. D. Hall,[1] G. Hammond,[36] M. Haney,[99] M. M. Hanke,[8] J. Hanks,[37] C. Hanna,[72] M. D. Hannam,[91] J. Hanson,[6] T. Hardwick,[2] J. Harms,[57,58] G. M. Harry,[101] I. W. Harry,[29] M. J. Hart,[36] M. T. Hartman,[5] C.-J. Haster,[45] K. Haughian,[36] J. Healy,[102] J. Heefner,[1,a] A. Heidmann,[60] M. C. Heintze,[5,6] G. Heinzel,[8] H. Heitmann,[53] P. Hello,[23] G. Hemming,[34] M. Hendry,[36] I. S. Heng,[36] J. Hennig,[36] A. W. Heptonstall,[1] M. Heurs,[8,17] S. Hild,[36] D. Hoak,[103] K. A. Hodge,[1] D. Hofman,[65] S. E. Hollitt,[104] K. Holt,[6] D. E. Holz,[75] P. Hopkins,[91] D. J. Hosken,[104] J. Hough,[36] E. A. Houston,[36] E. J. Howell,[51] Y. M. Hu,[36] S. Huang,[73] E. A. Huerta,[105,82] D. Huet,[23] B. Hughey,[97] S. Husa,[66] S. H. Huttner,[36] T. Huynh-Dinh,[6] A. Idrisy,[72] N. Indik,[8] D. R. Ingram,[37] R. Inta,[71] H. N. Isa,[36] J.-M. Isac,[60] M. Isi,[1] G. Islas,[22] T. Isogai,[10] B. R. Iyer,[15] K. Izumi,[37] M. B. Jacobson,[1] T. Jacqmin,[60] H. Jang,[77] K. Jani,[63] P. Jaranowski,[106] S. Jawahar,[107] F. Jiménez-Forteza,[66] W. W. Johnson,[2] N. K. Johnson-McDaniel,[15] D. I. Jones,[26] R. Jones,[36] R. J. G. Jonker,[9] L. Ju,[51] K. Haris,[108] C. V. Kalaghatgi,[24,91] V. Kalogera,[82] S. Kandhasamy,[21] G. Kang,[77] J. B. Kanner,[1] S. Karki,[59] M. Kasprzack,[2,23,34] E. Katsavounidis,[10] W. Katzman,[6] S. Kaufer,[17] T. Kaur,[51] K. Kawabe,[37] F. Kawazoe,[8,17] F. Kéfélian,[53] M. S. Kehl,[69] D. Keitel,[8,66] D. B. Kelley,[35] W. Kells,[1] R. Kennedy,[86] D. G. Keppel,[8] J. S. Key,[83] A. Khalaidovski,[8] F. Y. Khalili,[49] I. Khan,[12] S. Khan,[91] Z. Khan,[95] E. A. Khazanov,[109] N. Kijbunchoo,[37] C. Kim,[77] J. Kim,[110] K. Kim,[111] Nam-Gyu Kim,[77] Namjun Kim,[40] Y.-M. Kim,[110] E. J. King,[104] P. J. King,[37] D. L. Kinzel,[6] J. S. Kissel,[37] L. Kleybolte,[27] S. Klimenko,[5] S. M. Koehlenbeck,[8] K. Kokeyama,[2] S. Koley,[9] V. Kondrashov,[1] A. Kontos,[10] S. Koranda,[16] M. Korobko,[27] W. Z. Korth,[1] I. Kowalska,[44] D. B. Kozak,[1] V. Kringel,[8] B. Krishnan,[8] A. Królak,[112,113] C. Krueger,[17] G. Kuehn,[8] P. Kumar,[69] R. Kumar,[36] L. Kuo,[73] A. Kutynia,[112] P. Kwee,[8] B. D. Lackey,[35] M. Landry,[37] J. Lange,[102] B. Lantz,[40] P. D. Lasky,[114] A. Lazzarini,[1] C. Lazzaro,[63,42] P. Leaci,[29,79,28] S. Leavey,[36] E. O. Lebigot,[30,70] C. H. Lee,[110] H. K. Lee,[111] H. M. Lee,[115] K. Lee,[36] A. Lenon,[35] M. Leonardi,[89,90] J. R. Leong,[8] N. Leroy,[23] N. Letendre,[7] Y. Levin,[114] B. M. Levine,[37] T. G. F. Li,[1] A. Libson,[10] T. B. Littenberg,[116] N. A. Lockerbie,[107] J. Logue,[36] A. L. Lombardi,[103] L. T. London,[91] J. E. Lord,[35] M. Lorenzini,[12,13] V. Loriette,[117] M. Lormand,[6] G. Losurdo,[58] J. D. Lough,[8,17] C. O. Lousto,[102] G. Lovelace,[22] H. Lück,[17,8] A. P. Lundgren,[8] J. Luo,[78] R. Lynch,[10] Y. Ma,[51] T. MacDonald,[40] B. Machenschalk,[8] M. MacInnis,[10] D. M. Macleod,[2] F. Magaña-Sandoval,[35] R. M. Magee,[56] M. Mageswaran,[1] E. Majorana,[28] I. Maksimovic,[117] V. Malvezzi,[25,13] N. Man,[53] I. Mandel,[45] V. Mandic,[84] V. Mangano,[36] G. L. Mansell,[20] M. Manske,[16] M. Mantovani,[34] F. Marchesoni,[118,33] F. Marion,[7] S. Márka,[39] Z. Márka,[39] A. S. Markosyan,[40] E. Maros,[1] F. Martelli,[57,58] L. Martellini,[53] I. W. Martin,[36] R. M. Martin,[5] D. V. Martynov,[1] J. N. Marx,[1] K. Mason,[10] A. Masserot,[7] T. J. Massinger,[35] M. Masso-Reid,[36] F. Matichard,[10] L. Matone,[39] N. Mavalvala,[10] N. Mazumder,[56] G. Mazzolo,[8] R. McCarthy,[37] D. E. McClelland,[20] S. McCormick,[6] S. C. McGuire,[119] G. McIntyre,[1] J. McIver,[1] D. J. McManus,[20] S. T. McWilliams,[105] D. Meacher,[72] G. D. Meadors,[29,8] J. Meidam,[9] A. Melatos,[85] G. Mendell,[37] D. Mendoza-Gandara,[8] R. A. Mercer,[16] E. Merilh,[37] M. Merzougui,[53] S. Meshkov,[1] C. Messenger,[36] C. Messick,[72] P. M. Meyers,[84] F. Mezzani,[28,79] H. Miao,[45] C. Michel,[65] H. Middleton,[45] E. E. Mikhailov,[120] L. Milano,[67,4] J. Miller,[10] M. Millhouse,[31] Y. Minenkov,[13] J. Ming,[29,8] S. Mirshekari,[121] C. Mishra,[15] S. Mitra,[14] V. P. Mitrofanov,[49] G. Mitselmakher,[5] R. Mittleman,[10] A. Moggi,[19] M. Mohan,[34] S. R. P. Mohapatra,[10] M. Montani,[57,58] B. C. Moore,[88] C. J. Moore,[122] D. Moraru,[37] G. Moreno,[37] S. R. Morriss,[83] K. Mossavi,[8] B. Mours,[7] C. M. Mow-Lowry,[45] C. L. Mueller,[5] G. Mueller,[5] A. W. Muir,[91] Arunava Mukherjee,[15] D. Mukherjee,[16] S. Mukherjee,[83] N. Mukund,[14] A. Mullavey,[6] J. Munch,[104] D. J. Murphy,[39] P. G. Murray,[36] A. Mytidis,[5] I. Nardecchia,[25,13] L. Naticchioni,[79,28] R. K. Nayak,[123] V. Necula,[5] K. Nedkova,[103] G. Nelemans,[52,9] M. Neri,[46,47] A. Neunzert,[98] G. Newton,[36] T. T. Nguyen,[20] A. B. Nielsen,[8] S. Nissanke,[52,9] A. Nitz,[8] F. Nocera,[34] D. Nolting,[6] M. E. N. Normandin,[83] L. K. Nuttall,[35] J. Oberling,[37] E. Ochsner,[16] J. O'Dell,[100] E. Oelker,[10] G. H. Ogin,[124] J. J. Oh,[125] S. H. Oh,[125] F. Ohme,[91] M. Oliver,[66] P. Oppermann,[8] Richard J. Oram,[6] B. O'Reilly,[6] R. O'Shaughnessy,[102] C. D. Ott,[76] D. J. Ottaway,[104] R. S. Ottens,[5] H. Overmier,[6] B. J. Owen,[71] A. Pai,[108] S. A. Pai,[48] J. R. Palamos,[59] O. Palashov,[109] C. Palomba,[28] A. Pal-Singh,[27] H. Pan,[73] Y. Pan,[62] C. Pankow,[82] F. Pannarale,[91] B. C. Pant,[48] F. Paoletti,[34,19] A. Paoli,[34] M. A. Papa,[29,16,8] H. R. Paris,[40] W. Parker,[6] D. Pascucci,[36] A. Pasqualetti,[34] R. Passaquieti,[18,19] D. Passuello,[19] B. Patricelli,[18,19] Z. Patrick,[40] B. L. Pearlstone,[36] M. Pedraza,[1] R. Pedurand,[65] L. Pekowsky,[35] A. Pele,[6] S. Penn,[126] A. Perreca,[1] H. P. Pfeiffer,[69,29] M. Phelps,[36] O. Piccinni,[79,28] M. Pichot,[53] M. Pickenpack,[8] F. Piergiovanni,[57,58]

V. Pierro,[87] G. Pillant,[34] L. Pinard,[65] I. M. Pinto,[87] M. Pitkin,[36] J. H. Poeld,[8] R. Poggiani,[18,19] P. Popolizio,[34] A. Post,[8] J. Powell,[36] J. Prasad,[14] V. Predoi,[91] S. S. Premachandra,[114] T. Prestegard,[84] L. R. Price,[1] M. Prijatelj,[34] M. Principe,[87] S. Privitera,[29] R. Prix,[8] G. A. Prodi,[89,90] L. Prokhorov,[49] O. Puncken,[8] M. Punturo,[33] P. Puppo,[28] M. Pürrer,[29] H. Qi,[16] J. Qin,[51] V. Quetschke,[83] E. A. Quintero,[1] R. Quitzow-James,[59] F. J. Raab,[37] D. S. Rabeling,[20] H. Radkins,[37] P. Raffai,[54] S. Raja,[48] M. Rakhmanov,[83] C. R. Ramet,[6] P. Rapagnani,[79,28] V. Raymond,[29] M. Razzano,[18,19] V. Re,[25] J. Read,[22] C. M. Reed,[37] T. Regimbau,[53] L. Rei,[47] S. Reid,[50] D. H. Reitze,[1,5] H. Rew,[120] S. D. Reyes,[35] F. Ricci,[79,28] K. Riles,[98] N. A. Robertson,[1,36] R. Robie,[36] F. Robinet,[23] A. Rocchi,[13] L. Rolland,[7] J. G. Rollins,[1] V. J. Roma,[59] J. D. Romano,[83] R. Romano,[3,4] G. Romanov,[120] J. H. Romie,[6] D. Rosińska,[127,43] S. Rowan,[36] A. Rüdiger,[8] P. Ruggi,[34] K. Ryan,[37] S. Sachdev,[1] T. Sadecki,[37] L. Sadeghian,[16] L. Salconi,[34] M. Saleem,[108] F. Salemi,[8] A. Samajdar,[123] L. Sammut,[85,114] L. M. Sampson,[82] E. J. Sanchez,[1] V. Sandberg,[37] B. Sandeen,[82] G. H. Sanders,[1] J. R. Sanders,[98,35] B. Sassolas,[65] B. S. Sathyaprakash,[91] P. R. Saulson,[35] O. Sauter,[98] R. L. Savage,[37] A. Sawadsky,[17] P. Schale,[59] R. Schilling,[8,b] J. Schmidt,[8] P. Schmidt,[1,76] R. Schnabel,[27] R. M. S. Schofield,[59] A. Schönbeck,[27] E. Schreiber,[8] D. Schuette,[8,17] B. F. Schutz,[91,29] J. Scott,[36] S. M. Scott,[20] D. Sellers,[6] A. S. Sengupta,[94] D. Sentenac,[34] V. Sequino,[25,13] A. Sergeev,[109] G. Serna,[22] Y. Setyawati,[52,9] A. Sevigny,[37] D. A. Shaddock,[20] T. Shaffer,[37] S. Shah,[52,9] M. S. Shahriar,[82] M. Shaltev,[8] Z. Shao,[1] B. Shapiro,[40] P. Shawhan,[62] A. Sheperd,[16] D. H. Shoemaker,[10] D. M. Shoemaker,[63] K. Siellez,[53,63] X. Siemens,[16] D. Sigg,[37] A. D. Silva,[11] D. Simakov,[8] A. Singer,[1] L. P. Singer,[68] A. Singh,[29,8] R. Singh,[2] A. Singhal,[12] A. M. Sintes,[66] B. J. J. Slagmolen,[20] J. R. Smith,[22] M. R. Smith,[1] N. D. Smith,[1] R. J. E. Smith,[1] E. J. Son,[125] B. Sorazu,[36] F. Sorrentino,[47] T. Souradeep,[14] A. K. Srivastava,[95] A. Staley,[39] M. Steinke,[8] J. Steinlechner,[36] S. Steinlechner,[36] D. Steinmeyer,[8,17] B. C. Stephens,[16] S. P. Stevenson,[45] R. Stone,[83] K. A. Strain,[36] N. Straniero,[65] G. Stratta,[57,58] N. A. Strauss,[78] S. Strigin,[49] R. Sturani,[121] A. L. Stuver,[6] T. Z. Summerscales,[128] L. Sun,[85] P. J. Sutton,[91] B. L. Swinkels,[34] M. J. Szczepańczyk,[97] M. Tacca,[30] D. Talukder,[59] D. B. Tanner,[5] M. Tápai,[96] S. P. Tarabrin,[8] A. Taracchini,[29] R. Taylor,[1] T. Theeg,[8] M. P. Thirugnanasambandam,[1] E. G. Thomas,[45] M. Thomas,[6] P. Thomas,[37] K. A. Thorne,[6] K. S. Thorne,[76] E. Thrane,[114] S. Tiwari,[12] V. Tiwari,[91] K. V. Tokmakov,[107] C. Tomlinson,[86] M. Tonelli,[18,19] C. V. Torres,[83,c] C. I. Torrie,[1] D. Töyrä,[45] F. Travasso,[32,33] G. Traylor,[6] D. Trifirò,[21] M. C. Tringali,[89,90] L. Trozzo,[129,19] M. Tse,[10] M. Turconi,[53] D. Tuyenbayev,[83] D. Ugolini,[130] C. S. Unnikrishnan,[99] A. L. Urban,[16] S. A. Usman,[35] H. Vahlbruch,[17] G. Vajente,[1] G. Valdes,[83] M. Vallisneri,[76] N. van Bakel,[9] M. van Beuzekom,[9] J. F. J. van den Brand,[61,9] C. Van Den Broeck,[9] D. C. Vander-Hyde,[35,22] L. van der Schaaf,[9] J. V. van Heijningen,[9] A. A. van Veggel,[36] M. Vardaro,[41,42] S. Vass,[1] M. Vasúth,[38] R. Vaulin,[10] A. Vecchio,[45] G. Vedovato,[42] J. Veitch,[45] P. J. Veitch,[104] K. Venkateswara,[131] D. Verkindt,[7] F. Vetrano,[57,58] A. Viceré,[57,58] S. Vinciguerra,[45] D. J. Vine,[50] J.-Y. Vinet,[53] S. Vitale,[10] T. Vo,[35] H. Vocca,[32,33] C. Vorvick,[37] D. Voss,[5] W. D. Vousden,[45] S. P. Vyatchanin,[49] A. R. Wade,[20] L. E. Wade,[132] M. Wade,[132] S. J. Waldman,[10] M. Walker,[2] L. Wallace,[1] S. Walsh,[16,8,29] G. Wang,[12] H. Wang,[45] M. Wang,[45] X. Wang,[70] Y. Wang,[51] H. Ward,[36] R. L. Ward,[20] J. Warner,[37] M. Was,[7] B. Weaver,[37] L.-W. Wei,[53] M. Weinert,[8] A. J. Weinstein,[1] R. Weiss,[10] T. Welborn,[6] L. Wen,[51] P. Weßels,[8] T. Westphal,[8] K. Wette,[8] J. T. Whelan,[102,8] S. E. Whitcomb,[1] D. J. White,[86] B. F. Whiting,[5] K. Wiesner,[8] C. Wilkinson,[37] P. A. Willems,[1] L. Williams,[5] R. D. Williams,[1] A. R. Williamson,[91] J. L. Willis,[133] B. Willke,[17,8] M. H. Wimmer,[8,17] L. Winkelmann,[8] W. Winkler,[8] C. C. Wipf,[1] A. G. Wiseman,[16] H. Wittel,[8,17] G. Woan,[36] J. Worden,[37] J. L. Wright,[36] G. Wu,[6] J. Yablon,[82] I. Yakushin,[6] W. Yam,[10] H. Yamamoto,[1] C. C. Yancey,[62] M. J. Yap,[20] H. Yu,[10] M. Yvert,[7] A. Zadrożny,[112] L. Zangrando,[42] M. Zanolin,[97] J.-P. Zendri,[42] M. Zevin,[82] F. Zhang,[10] L. Zhang,[1] M. Zhang,[120] Y. Zhang,[102] C. Zhao,[51] M. Zhou,[82] Z. Zhou,[82] X. J. Zhu,[51] M. E. Zucker,[1,10] S. E. Zuraw,[103] and J. Zweizig[1]

(LIGO Scientific Collaboration and Virgo Collaboration)

[1]*LIGO, California Institute of Technology, Pasadena, California 91125, USA*
[2]*Louisiana State University, Baton Rouge, Louisiana 70803, USA*
[3]*Università di Salerno, Fisciano, I-84084 Salerno, Italy*
[4]*INFN, Sezione di Napoli, Complesso Universitario di Monte S. Angelo, I-80126 Napoli, Italy*
[5]*University of Florida, Gainesville, Florida 32611, USA*
[6]*LIGO Livingston Observatory, Livingston, Louisiana 70754, USA*
[7]*Laboratoire d'Annecy-le-Vieux de Physique des Particules (LAPP), Université Savoie Mont Blanc, CNRS/IN2P3, F-74941 Annecy-le-Vieux, France*
[8]*Albert-Einstein-Institut, Max-Planck-Institut für Gravitationsphysik, D-30167 Hannover, Germany*
[9]*Nikhef, Science Park, 1098 XG Amsterdam, Netherlands*

[10]LIGO, Massachusetts Institute of Technology, Cambridge, Massachusetts 02139, USA
[11]Instituto Nacional de Pesquisas Espaciais, 12227-010 São José dos Campos, São Paulo, Brazil
[12]INFN, Gran Sasso Science Institute, I-67100 L'Aquila, Italy
[13]INFN, Sezione di Roma Tor Vergata, I-00133 Roma, Italy
[14]Inter-University Centre for Astronomy and Astrophysics, Pune 411007, India
[15]International Centre for Theoretical Sciences, Tata Institute of Fundamental Research, Bangalore 560012, India
[16]University of Wisconsin-Milwaukee, Milwaukee, Wisconsin 53201, USA
[17]Leibniz Universität Hannover, D-30167 Hannover, Germany
[18]Università di Pisa, I-56127 Pisa, Italy
[19]INFN, Sezione di Pisa, I-56127 Pisa, Italy
[20]Australian National University, Canberra, Australian Capital Territory 0200, Australia
[21]The University of Mississippi, University, Mississippi 38677, USA
[22]California State University Fullerton, Fullerton, California 92831, USA
[23]LAL, Université Paris-Sud, CNRS/IN2P3, Université Paris-Saclay, Orsay, France
[24]Chennai Mathematical Institute, Chennai, India 603103
[25]Università di Roma Tor Vergata, I-00133 Roma, Italy
[26]University of Southampton, Southampton SO17 1BJ, United Kingdom
[27]Universität Hamburg, D-22761 Hamburg, Germany
[28]INFN, Sezione di Roma, I-00185 Roma, Italy
[29]Albert-Einstein-Institut, Max-Planck-Institut für Gravitationsphysik, D-14476 Potsdam-Golm, Germany
[30]APC, AstroParticule et Cosmologie, Université Paris Diderot, CNRS/IN2P3, CEA/Irfu, Observatoire de Paris, Sorbonne Paris Cité, F-75205 Paris Cedex 13, France
[31]Montana State University, Bozeman, Montana 59717, USA
[32]Università di Perugia, I-06123 Perugia, Italy
[33]INFN, Sezione di Perugia, I-06123 Perugia, Italy
[34]European Gravitational Observatory (EGO), I-56021 Cascina, Pisa, Italy
[35]Syracuse University, Syracuse, New York 13244, USA
[36]SUPA, University of Glasgow, Glasgow G12 8QQ, United Kingdom
[37]LIGO Hanford Observatory, Richland, Washington 99352, USA
[38]Wigner RCP, RMKI, H-1121 Budapest, Konkoly Thege Miklós út 29-33, Hungary
[39]Columbia University, New York, New York 10027, USA
[40]Stanford University, Stanford, California 94305, USA
[41]Università di Padova, Dipartimento di Fisica e Astronomia, I-35131 Padova, Italy
[42]INFN, Sezione di Padova, I-35131 Padova, Italy
[43]CAMK-PAN, 00-716 Warsaw, Poland
[44]Astronomical Observatory Warsaw University, 00-478 Warsaw, Poland
[45]University of Birmingham, Birmingham B15 2TT, United Kingdom
[46]Università degli Studi di Genova, I-16146 Genova, Italy
[47]INFN, Sezione di Genova, I-16146 Genova, Italy
[48]RRCAT, Indore MP 452013, India
[49]Faculty of Physics, Lomonosov Moscow State University, Moscow 119991, Russia
[50]SUPA, University of the West of Scotland, Paisley PA1 2BE, United Kingdom
[51]University of Western Australia, Crawley, Western Australia 6009, Australia
[52]Department of Astrophysics/IMAPP, Radboud University Nijmegen, P.O. Box 9010, 6500 GL Nijmegen, Netherlands
[53]Artemis, Université Côte d'Azur, CNRS, Observatoire Côte d'Azur, CS 34229, Nice cedex 4, France
[54]MTA Eötvös University, "Lendulet" Astrophysics Research Group, Budapest 1117, Hungary
[55]Institut de Physique de Rennes, CNRS, Université de Rennes 1, F-35042 Rennes, France
[56]Washington State University, Pullman, Washington 99164, USA
[57]Università degli Studi di Urbino "Carlo Bo," I-61029 Urbino, Italy
[58]INFN, Sezione di Firenze, I-50019 Sesto Fiorentino, Firenze, Italy
[59]University of Oregon, Eugene, Oregon 97403, USA
[60]Laboratoire Kastler Brossel, UPMC-Sorbonne Universités, CNRS, ENS-PSL Research University, Collège de France, F-75005 Paris, France
[61]VU University Amsterdam, 1081 HV Amsterdam, Netherlands
[62]University of Maryland, College Park, Maryland 20742, USA
[63]Center for Relativistic Astrophysics and School of Physics, Georgia Institute of Technology, Atlanta, Georgia 30332, USA
[64]Institut Lumière Matière, Université de Lyon, Université Claude Bernard Lyon 1, UMR CNRS 5306, 69622 Villeurbanne, France
[65]Laboratoire des Matériaux Avancés (LMA), IN2P3/CNRS, Université de Lyon, F-69622 Villeurbanne, Lyon, France
[66]Universitat de les Illes Balears, IAC3—IEEC, E-07122 Palma de Mallorca, Spain
[67]Università di Napoli "Federico II," Complesso Universitario di Monte S. Angelo, I-80126 Napoli, Italy

[68]*NASA/Goddard Space Flight Center, Greenbelt, Maryland 20771, USA*
[69]*Canadian Institute for Theoretical Astrophysics, University of Toronto, Toronto, Ontario M5S 3H8, Canada*
[70]*Tsinghua University, Beijing 100084, China*
[71]*Texas Tech University, Lubbock, Texas 79409, USA*
[72]*The Pennsylvania State University, University Park, Pennsylvania 16802, USA*
[73]*National Tsing Hua University, Hsinchu City, 30013 Taiwan, Republic of China*
[74]*Charles Sturt University, Wagga Wagga, New South Wales 2678, Australia*
[75]*University of Chicago, Chicago, Illinois 60637, USA*
[76]*Caltech CaRT, Pasadena, California 91125, USA*
[77]*Korea Institute of Science and Technology Information, Daejeon 305-806, Korea*
[78]*Carleton College, Northfield, Minnesota 55057, USA*
[79]*Università di Roma "La Sapienza," I-00185 Roma, Italy*
[80]*University of Brussels, Brussels 1050, Belgium*
[81]*Sonoma State University, Rohnert Park, California 94928, USA*
[82]*Northwestern University, Evanston, Illinois 60208, USA*
[83]*The University of Texas Rio Grande Valley, Brownsville, Texas 78520, USA*
[84]*University of Minnesota, Minneapolis, Minnesota 55455, USA*
[85]*The University of Melbourne, Parkville, Victoria 3010, Australia*
[86]*The University of Sheffield, Sheffield S10 2TN, United Kingdom*
[87]*University of Sannio at Benevento, I-82100 Benevento, Italy and INFN, Sezione di Napoli, I-80100 Napoli, Italy*
[88]*Montclair State University, Montclair, New Jersey 07043, USA*
[89]*Università di Trento, Dipartimento di Fisica, I-38123 Povo, Trento, Italy*
[90]*INFN, Trento Institute for Fundamental Physics and Applications, I-38123 Povo, Trento, Italy*
[91]*Cardiff University, Cardiff CF24 3AA, United Kingdom*
[92]*National Astronomical Observatory of Japan, 2-21-1 Osawa, Mitaka, Tokyo 181-8588, Japan*
[93]*School of Mathematics, University of Edinburgh, Edinburgh EH9 3FD, United Kingdom*
[94]*Indian Institute of Technology, Gandhinagar Ahmedabad Gujarat 382424, India*
[95]*Institute for Plasma Research, Bhat, Gandhinagar 382428, India*
[96]*University of Szeged, Dóm tér 9, Szeged 6720, Hungary*
[97]*Embry-Riddle Aeronautical University, Prescott, Arizona 86301, USA*
[98]*University of Michigan, Ann Arbor, Michigan 48109, USA*
[99]*Tata Institute of Fundamental Research, Mumbai 400005, India*
[100]*Rutherford Appleton Laboratory, HSIC, Chilton, Didcot, Oxon OX11 0QX, United Kingdom*
[101]*American University, Washington, D.C. 20016, USA*
[102]*Rochester Institute of Technology, Rochester, New York 14623, USA*
[103]*University of Massachusetts-Amherst, Amherst, Massachusetts 01003, USA*
[104]*University of Adelaide, Adelaide, South Australia 5005, Australia*
[105]*West Virginia University, Morgantown, West Virginia 26506, USA*
[106]*University of Biał ystok, 15-424 Biał ystok, Poland*
[107]*SUPA, University of Strathclyde, Glasgow G1 1XQ, United Kingdom*
[108]*IISER-TVM, CET Campus, Trivandrum Kerala 695016, India*
[109]*Institute of Applied Physics, Nizhny Novgorod, 603950, Russia*
[110]*Pusan National University, Busan 609-735, Korea*
[111]*Hanyang University, Seoul 133-791, Korea*
[112]*NCBJ, 05-400 Świerk-Otwock, Poland*
[113]*IM-PAN, 00-956 Warsaw, Poland*
[114]*Monash University, Victoria 3800, Australia*
[115]*Seoul National University, Seoul 151-742, Korea*
[116]*University of Alabama in Huntsville, Huntsville, Alabama 35899, USA*
[117]*ESPCI, CNRS, F-75005 Paris, France*
[118]*Università di Camerino, Dipartimento di Fisica, I-62032 Camerino, Italy*
[119]*Southern University and A&M College, Baton Rouge, Louisiana 70813, USA*
[120]*College of William and Mary, Williamsburg, Virginia 23187, USA*
[121]*Instituto de Física Teórica, University Estadual Paulista/ICTP South American Institute for Fundamental Research, São Paulo SP 01140-070, Brazil*
[122]*University of Cambridge, Cambridge CB2 1TN, United Kingdom*
[123]*IISER-Kolkata, Mohanpur, West Bengal 741252, India*
[124]*Whitman College, 345 Boyer Avenue, Walla Walla, Washington 99362 USA*
[125]*National Institute for Mathematical Sciences, Daejeon 305-390, Korea*
[126]*Hobart and William Smith Colleges, Geneva, New York 14456, USA*

[127]*Janusz Gil Institute of Astronomy, University of Zielona Góra, 65-265 Zielona Góra, Poland*
[128]*Andrews University, Berrien Springs, Michigan 49104, USA*
[129]*Università di Siena, I-53100 Siena, Italy*
[130]*Trinity University, San Antonio, Texas 78212, USA*
[131]*University of Washington, Seattle, Washington 98195, USA*
[132]*Kenyon College, Gambier, Ohio 43022, USA*
[133]*Abilene Christian University, Abilene, Texas 79699, USA*

[a]Deceased, April 2012.
[b]Deceased, May 2015.
[c]Deceased, March 2015.